\documentstyle[11pt,epsf]{article}
\makeatletter
\long\def\@makecaption#1#2{%
  \vskip\abovecaptionskip
  \sbox\@tempboxa{\footnotesize #1: #2}%
  \ifdim \wd\@tempboxa >\hsize
    \footnotesize #1: #2\par
  \else
    \global \@minipagefalse
    \hbox to\hsize{\hfil\box\@tempboxa\hfil}%
  \fi
  \vskip\belowcaptionskip}
\makeatother

\newcommand{\Tr}{\mathop{\rm Tr}}
\newcommand{\bbox}[1]{\mbox{\boldmath $#1$}}
\begin{document}
\title{
\vspace{-60pt}
\rightline{\large\sf KUNS1463}
\vspace{12pt}
Periodic-Orbit Bifurcation and Shell Structure at Exotic
Deformation\thanks{
Talk presented by K.M. at the International Symposium on
{\it ``Atomic Nuclei and Metallic Clusters: Finite Many-Fermion Systems''},
Prague, Sep. 1--5, 1997.}
}
\author{Ken-ichiro Arita, Ayumu Sugita$^\dagger$
and Kenichi Matsuyanagi$^\dagger$\\
\normalsize
\it Department of Physics, Nagoya Institute of Technology,
\it Nagoya 466, Japan\\
\it $^\dagger$Department of Physics, Graduate School of Science, Kyoto
University,\\
\it Kyoto 606-01, Japan}
\date{}
\maketitle

\begin{abstract}
By means of periodic orbit theory and deformed cavity model, we have
investigated semiclassical origin of superdeformed shell structure and
also of reflection-asymmetric deformed shapes.  Systematic analysis of
quantum-classical correspondence reveals that bifurcation of
equatorial orbits into three-dimensional ones play predominant role in
the formation of these shell structures.

\vspace*{5pt}
\end{abstract}

\section{Introduction}

Shell structure, i.e., regular oscillating pattern in the smoothed
single-particle level density, coarse-grained with respect to energy
resolution, plays decisive role in determining shapes of finite
Fermion systems
[1--6].  According to the periodic-orbit theory
[7--11] based on the semiclassical approximation to the path integral,
shell structure is determined by classical periodic orbits with short
periods.  Finite Fermion systems like nuclei and metallic clusters
favor such shapes at which prominent shell structures are formed and
their Fermi surfaces lie in the valley of the oscillating level
density, increasing their binding energies in this manner.

In this talk, we investigate the axially-symmetric deformed cavity
model as a simple model of single-particle motions in nuclei and
metallic clusters \cite{balian,stru,frisk}, and try to find the
correspondence between quantum shell structure and classical periodic
orbits.  Our major purpose is to identify most important periodic
orbits that determine major patterns of oscillating level densities at
exotic deformations including prolate superdeformations, prolate
hyperdeformations, oblate superdeformations and reflection-asymmetric
shapes.

In the cavity model, the action integral $S_{\gamma}$ for a periodic
orbit $\gamma$ is proportional to the length $L_{\gamma}$ of it,
$S_{\gamma}=\oint_{\gamma} \bbox{p}\cdot d\bbox{q}=\hbar kL_{\gamma}$,
and the trace formula for the oscillating part of the level density is
written as
\begin{equation}
\tilde{\rho}(E)\simeq\sum_{\gamma}A_{\gamma}
k^{(d_{\gamma}-2)/2}\cos(kL_{\gamma}-\pi\mu_\gamma/2),
\label{eq:tracef}
\end{equation}
where $d_{\gamma}$ and $\mu_{\gamma}$ denote the degeneracy and the
Maslov phase of the periodic orbit $\gamma$, respectively.  Fourier
transform $\tilde{F}(L)$ of $\tilde{\rho}(E)$ with respect to wave
number $k$ is

\begin{eqnarray}
\tilde{F}(L)&=&\int dk\, k^{-(d-2)/2}e^{-ikL}
    \tilde{\rho}(E=\hbar^2k^2/2M) \nonumber\\
    &\simeq& \sum_{\gamma}A'_{\gamma}\delta(L-L_{\gamma}),
\end{eqnarray}
which may be regarded as `length spectrum' exhibiting peaks at lengths
of individual periodic orbits.  In the following, we shall make full
use of the Fourier transforms in order to identify important periodic
orbits.

We solve the Schr\"odinger equation for single-particle motions in the
cavity under the Dirichlet boundary condition.  We have constructed a
computer program by which we can efficiently obtain a large number of
eigenvalues as function of deformation parameters of the cavity
\cite{SAM}.
We have systematically searched for classical periodic orbits in the
three-dimensional(3D) cavities on the basis of the monodromy method
\cite{baranger}.

\section{Periodic-orbit bifurcations}

As is well known, only linear and planar orbits exist in the spherical
limit.  When quadrupole deformation sets in, linear (diameter) orbits
bifurcate into those along the major axis and along the minor axis.
Likewise, planar orbits bifurcate into those in the meridian plane
(containing the symmetry axis) and in the equatorial plane
(perpendicular to the symmetry axis).

With variation of deformation, 3D and new 2D periodic orbits are 
successively born through bifurcations.  Bifurcations that are
important in the following discussions are\\
(i) bifurcations from multiple repetitions along the minor axis,
which generate butterfly-shaped planar orbits in the meridian
plane, and\\
(ii) bifurcations from multiple traversals of planar orbits in 
the equatorial plane, which generate 3D periodic orbits.

For prolate shapes (i) may be regarded as a limit of (ii), while this
distinction is important for oblate shapes.  We shall see that
bifurcations of type (ii) are especially important for shell structure
at prolate super- and hyper-deformations and at reflection-asymmetric
shapes.  Bifurcation points for (ii) are determined by stability of
equatorial-plane orbits against small displacements in the
longitudinal direction. Bifurcations occur when the following
condition is met:
\begin{equation}
\frac{R_2}{R_1}=\frac{\sin(\pi t/p)^2}{\sin(\pi q/p)^2},
\label{eq:bifcond}
\end{equation}     
where $R_1$ and $R_2$ denote the main curvature radii for the
longitudinal and equatorial directions, respectively, and
($p$,$t$,$q$) are positive integers.

At the bifurcation points, trace of the $(2\times 2)$ reduced
monodromy matrix $\bbox{M}$ representing stabilities of
equatorial-plane orbits becomes $\Tr\bbox{M}=2$, indicating that they
are of neutral stability at these points.  The above equation was first
derived by Balian and Bloch \cite{balian}.  We note that, for the
special case of prolate spheroidal shapes, $R_2/R_1$ is simply related
to the axis ratio $a/b$ as $R_2/R_1=(a/b)^2$, $a$ and $b$ being
lengths of the major and the minor axes, respectively, and
($p$,$t$,$q$) represent the numbers of vibrations or rotations of the
periodic orbits with respect to the three spheroidal coordinates.
They correspond to $(n_{\epsilon},n_{\phi},n_{\xi})$ and
$(n_v,n_{\phi},n_u)$ of Refs.~\cite{stru,nishi1}, respectively.
Periodic-orbit bifurcations in spheroidal cavities have been
thoroughly studied by Nishioka et al.\cite{nishi1,nishi2}

\section{Semiclassical origin of superdeformations}

\begin{figure}[t]
\epsfxsize=.75\textwidth
\centerline{\epsffile{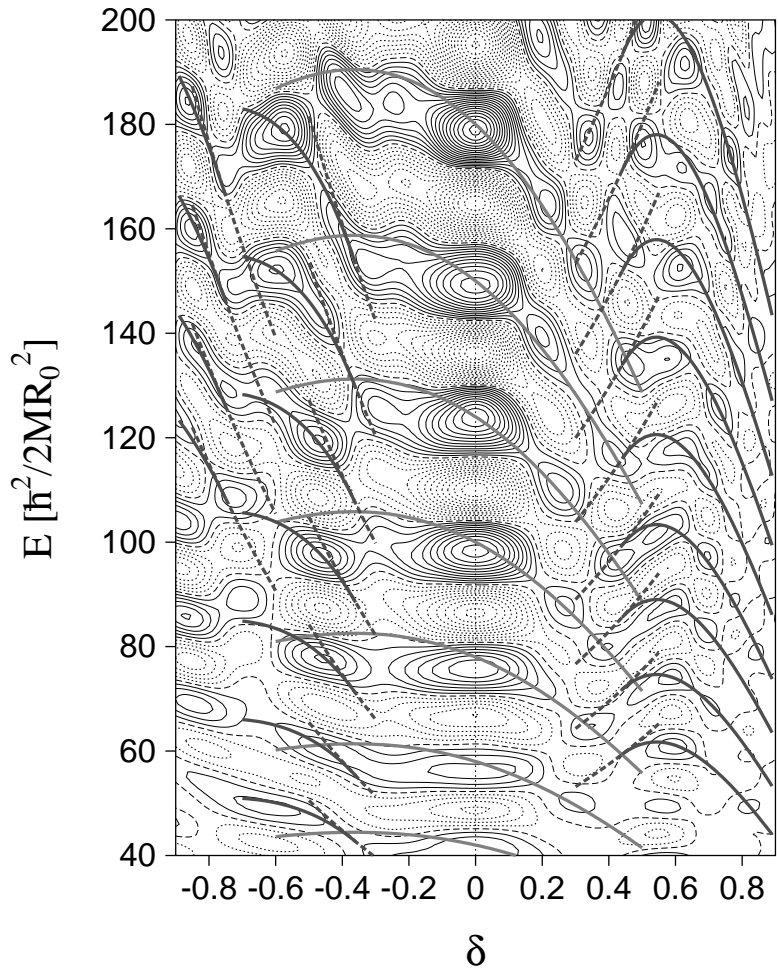}}
\caption{\label{fig:sldmap}
Oscillating parts of the smoothed level densities for spheroidal
cavities, displayed as function of energy (in unit of
$\hbar^2/2MR_0^2$) and deformation.  Constant-action lines for some
short periodic orbits are indicated by thick solid and broken lines
(see text).  The deformation parameter $\delta$ is related to the axis
ratio $\eta\equiv a/b$ by $\delta=3(\eta-1)/(2\eta+1)$ in the prolate
case and by $\delta=-3(\eta-1)/(\eta+2)$ in the oblate case.}
\end{figure}

Let us first discuss spheroidal cavities.  In Fig.~\ref{fig:sldmap}
oscillating parts of the smoothed level densities are displayed in a
form of contour map with respect to energy and deformation. Regular
patterns consisting of several valley-ridge structures are clearly
seen.  Thick solid and broken lines indicate constant-action lines for
some important periodic orbits on which we are going to discuss.  We
here note that, as emphasized by Strutinsky et al.\cite{stru}, if few
families of orbits having almost the same values of action integral
$S_\gamma$ dominate in the sum in Eq.~(\ref{eq:tracef}), the valleys
in the contour map may follow such lines along which $S_\gamma$ stay
approximately constant.

\begin{figure}[t]
\epsfxsize=\textwidth
\centerline{\epsffile{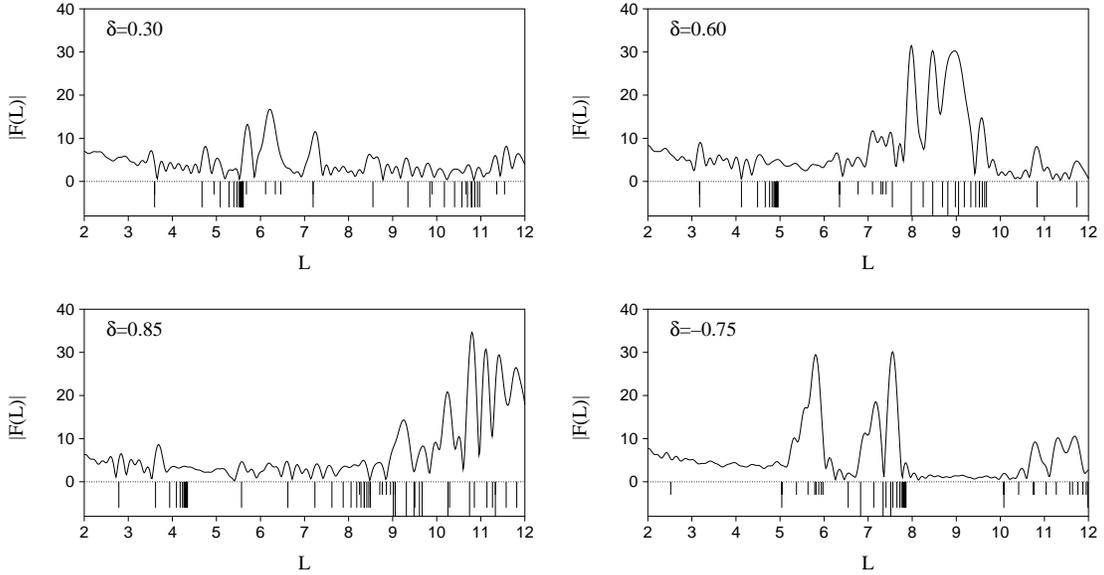}}
\caption{\label{fig:ftl}
Fourier transforms of quantum level densities for spheroidal cavities
with $\delta=0.3$, 0.6 (prolate superdeformation), 0.85 (prolate
hyperdeformation) and $-0.75$ (oblate superdeformation).  In the
bottoms of every figures, lengths (in unit of $R_0$) of classical
periodic orbits are indicated by vertical lines.  Long, middle and
short vertical lines are used for 3D orbits, planar orbits in the
equatorial and the meridian planes, respectively.}
\end{figure}

Figure~\ref{fig:ftl} displays Fourier transforms of the level
densities.  At normal deformation with $\delta=0.3$, we notice peaks
associated with triangular and quadrilateral orbits in the meridian
plane.

Constant-action lines for the triangular orbits are indicated in
Fig.~\ref{fig:sldmap} for several values of $e_F$ that go through the
spherical closed shells.  It is clear that the valleys run along these
lines.

With increasing deformation, bifurcations of linear and planar orbits
in the equatorial plane successively take place \cite{nishi1}:
Butterfly-shaped planar orbits with ($p$:$t$:$q$)=(4:2:1) bifurcate at
$\delta\simeq 0.32$ from double repetitions of linear orbits along the
minor axis. Then, 3D orbits (5:2:1) bifurcate at $\delta\simeq 0.44$
from five-point star-shaped orbits in the equatorial plane.  Similar
3D orbits (6:2:1), (7:2:1), (8:2:1), etc. successively bifurcate from
double traversals of triangular orbits, 7-point star-shaped orbits,
double traversals of rectangular orbits, etc. in the equatorial plane.
These 3D orbits form the prominent peaks seen in the range $L=8\sim9$
in the Fourier transform for $\delta=0.6$ (axis ratio 2:1).

Constant-action lines for the 3D orbits (5:2:1) are indicated by thick
solid lines in the region $\delta\geq 0.44$ of Fig.~\ref{fig:sldmap}.
Good correspondence is found between these lines and shapes of the
valleys seen in the superdeformed region.  Constant-action lines for
the other 3D orbits mentioned above also behave in the same fashion.

\begin{figure}
\epsfxsize=.8\textwidth
\centerline{\epsffile{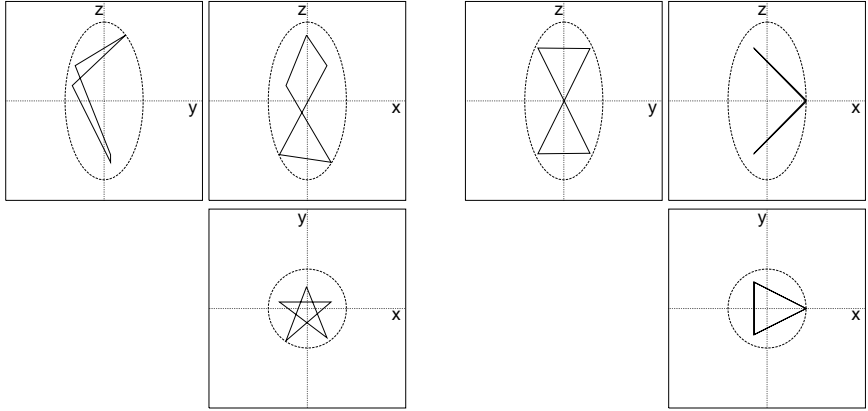}}
\caption{\label{fig:orbit3d}
Three-dimensional orbits (5:2:1) and (6:2:1) in the superdeformed
prolate cavity.  Their projections on the $(x,y)$, $(y,z)$ and $(z,x)$
planes are displayed.}
\end{figure}

Some of these 3D orbits are displayed in Fig.~\ref{fig:orbit3d}.  They
possess similarities with figure-eight shaped orbits in the
superdeformed harmonic oscillator with frequency ratio
$\omega_\perp$:$\omega_z$=2:1.  An important difference between the
the cavity model under consideration and the harmonic oscillator model
should be noted, however: In the former they exist for all deformation
parameters $\delta$ larger than the bifurcation points, whereas in the
latter such periodic orbits appear only for special deformations
corresponding to rational ratios of the major and the minor axes.%
\footnote{Note, however, that periodic orbits appear through
bifurcations also for irrational ratios, if anharmonic terms like octupole
deformations are added; see \cite{AM}.}

On the other hand, Fourier peak heights associated with new orbits
created by bifurcations quickly increase with increasing deformation
and reach the maxima.  Then, they start to decline.  Thus, with
variation of deformation, they are replaced by different periodic
orbits bifurcated later. We can confirm this, for instance, by
comparing the Fourier transform for $\delta=0.6$ (axis ratio 2:1) with
that for $\delta=0.85$ (axis ratio 3:1).  In the latter, we see
prominent peaks in the region $L=11\sim 12$ associated with 3D orbits
(7:3:1), (8:3:1), (9:3:1) that are bifurcated, respectively, at
$\delta\simeq 0.68, 0.73, 0,76$ from 7-point, 8-point star-shaped
orbits, and triple traversals of the triangular orbits in the
equatorial plane.  These 3D orbits resemble with Lissajous figures of
the hyperdeformed harmonic oscillator with the frequency ratio 3:1.

For oblate spheroidal cavities with $\delta=-0.75$ (axis ratio 1:2),
we see prominent peaks at $L\simeq 5.8$ associated with
butterfly-shaped planar orbits (4:1:1),
which are bifurcated at $\delta\simeq -0.36$ (axis ratio 1:$\sqrt 2$)
from double repetitions of linear orbits along the minor axis.  In
addition, just at this shape, new planar orbits (6:1:1) bifurcate from
triple repetitions of linear orbits along the minor axis
\cite{nishi2}.  We indeed see that a new peak associated with this
bifurcation arises at $L\simeq 7.6$.

Constant-action lines for these bifurcated orbits (4:1:1) and (6:1:1)
are indicated by thick solid lines in the region $\delta \leq -0.36$
of Fig.~\ref{fig:sldmap}.  We see clear correspondence between shapes
of these lines and of valleys in the oscillating level density.
Combining this good correspondence with the behavior of the Fourier
peaks mentioned above, it is evident that these periodic orbits are
responsible for the shell structure at oblate superdeformation.


The spheroidal cavities are special in that every bifurcated orbits
form continuous families of degeneracy two, which means that we need
two parameters to specify a single orbit among continuous set of
orbits belonging to a family having a common value of action integral
(length).  We have checked \cite{ASM}, however, that the results
obtained for spheroidal cavities persist also for other
parameterizations of quadrupole shapes where the degeneracy is one.
The present results for prolate normal- and super-deformations confirm
the qualitative argument by Strutinsky et al.\cite{stru}, except for
the strong deformation dependence, found above, of relative
contributions of different periodic orbits.
 
\begin{table}
\caption{\label{table}
List of bifurcation points of important periodic orbits in the
spheroidal cavity model.  For more details, see Nishioka et al.
\protect\cite{nishi1,nishi2}}
\begin{center}
\small
\catcode`?=\active \def?{\phantom{0}}
\begin{tabular}{cccc} \hline
{orbit ($p$:$t$:$q$)} &{axis ratio  $(a/b)$} &{deformation $\delta$}
&{orbit length in $R_0$}  \\ \hline
(4:2:1) & 1.41 &           0.32 & ?7.1 \\
(5:2:1) & 1.62 &           0.44 & ?8.1 \\
(6:2:1) & 1.73 &           0.49 & ?8.7 \\
(7:2:1) & 1.80 &           0.52 & ?9.0 \\
(8:2:1) & 1.85 &           0.54 & ?9.2 \\ \hline
(6:3:1) & 2.0? &           0.6? & ?9.5 \\
(7:3:1) & 2.26 &           0.68 & 10.3 \\
(8:3:1) & 2.42 &           0.73 & 10.9 \\
(9:3:1) & 2.53 &           0.76 & 11.4 \\ \hline
(4:1:1) & 1.41 & \llap{$-$}0.36 & ?6.4 \\
(6:1:1) & 2.0? & \llap{$-$}0.73 & ?7.6 \\ \hline 
\end{tabular}
\end{center}
\end{table}

\section{Reflection-asymmetric shapes}

To explore the possibilities that significant shell structures emerge
in the single-particle spectra for non-integrable Hamiltonians, we
have carried out analysis of single-particle motions in
reflection-asymmetric cavities by parameterizing the surface as
\begin{equation}
 R(\theta)/R_0=\frac 1{\sqrt{ (\frac {\cos\theta}a)^2 +
                                   (\frac {\sin\theta}b)^2 }} 
  + a_3 Y_{30}(\theta),
\end{equation}

\begin{figure}[b]
\epsfxsize=.8\textwidth
\centerline{\epsffile{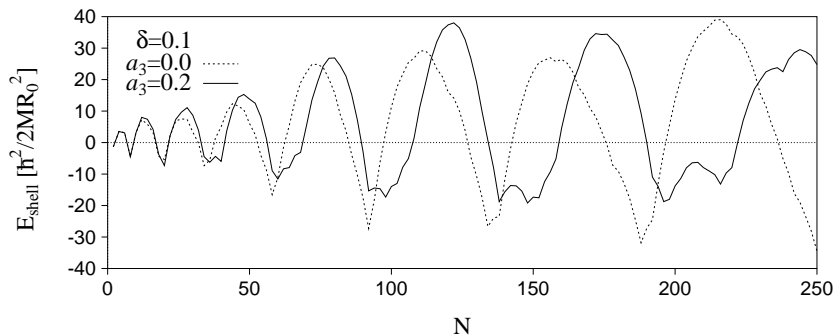}}
\caption{\label{fig:sce}
Shell structure energies (in unit of $\hbar^2/2MR_0^2$) of the
reflection-asymmetric cavity with $\delta=0.1$ and $a_3=0.2$,
evaluated with the Strutinsky method and plotted as function of
particle number $N$ counting the spin degeneracy factor of two.  For
comparison, those for $\delta=0.1$ and $a_3=0.0$ are plotted by broken
lines.}
\end{figure}

When octupole deformation is added to the prolate shape (at normal
deformation), spheroidal symmetry is broken and, accordingly,
contribution of the triangular and quadrilateral orbits in the
meridian plane decline.  However, we have found that remarkable shell
structure emerges for certain combinations of quadrupole and octupole
deformations \cite{AM,SAM}.  As an example, Fig.~\ref{fig:sce} shows
shell-structure energies calculated for $\delta=0.1$ and $a_3=0.2$
with the Strutinsky procedure.  Remarkable shell-energy gains are
obtained by such deformations for systems above the spherical closed
shells. This appears consistent with the result of realistic
calculations by Frauendorf and Pashkevich \cite{FP} for shapes of
sodium clusters.

\begin{figure}
\epsfxsize=.6\textwidth
\centerline{\epsffile{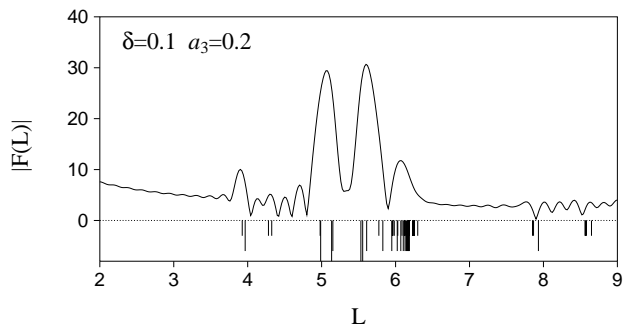}}
\caption{\label{fig:ftl2}
Same as Fig.~\protect\ref{fig:ftl}, but for reflection-asymmetric
cavity with $\delta=0.1$ and $a_3=0.2$.}
\end{figure}

Semiclassical origin of this quadrupole-octupole shell structure is
again connected with bifurcation of `equatorial'-plane orbits.
Figure~\ref{fig:ftl2} shows the Fourier transform.  We can clearly
identify new peaks associated with orbits (3:1:1) and (4:1:1)
bifurcated from triangular and square orbits in the `equatorial' plane
at the center of the larger cluster of the pear-shaped cavity.

\begin{figure}[t]
\epsfxsize=.2\textwidth
\centerline{\epsffile{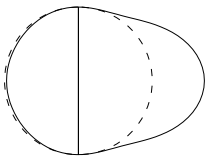}}
\caption{\label{fig:curv}
Illustration of a shape at the bifurcation point.  A sphere tangent to
the boundary at the 'equatorial' plane is indicated by a broken line.}
\end{figure}

The key to understand the reason why bifurcations from
`equatorial'-plane orbits play important roles at finite octupole
deformations may lie in the following point: Stability of these orbits
is crucially dependent on the curvature of the boundary.  The
curvature radius in the longitudinal direction changes as the octupole
deformation parameter $a_3$ varies, and at certain combinations of
$\delta$ and $a_3$, it matches with the equatorial radius, as
illustrated in Fig.~\ref{fig:curv}.  At this point, periodic orbits
in the equatorial plane acquire {\em local spherical symmetry}, and
form local continuous set of periodic orbits leaving from the
`equatorial' plane.  This continuous set makes a coherent contribution
to the trace integral and significantly enhances the amplitudes
associated with these orbits.  This is just the bifurcation point of
orbits in the `equatorial' plane, and 3D orbits bifurcate from the
above local continuous set.  One can readily check that for $R_2=R_1$
all orbits ($p=2,3,4,\ldots$) in the `equatorial' plane simultaneously
satisfy the bifurcation condition (\ref{eq:bifcond}) with $t=q=1$.

\begin{figure}[b]
\epsfxsize=.6\textwidth
\centerline{\epsffile{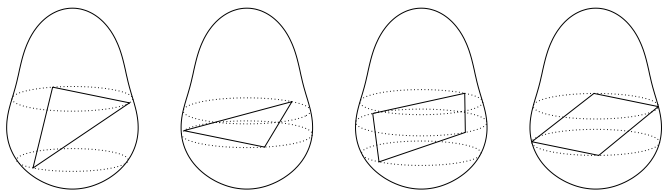}}
\caption{\label{fig:orbit}
Some short periodic orbits bifurcated from `equatorial'-plane orbits.}
\end{figure}

Some periodic orbits born out of these bifurcations are displayed in
Fig.~\ref{fig:orbit}.  Note that octupole deformations play crucial
role in creating this kind of bifurcations, that occurs from a single
turn ($t=1$) of the `equatorial'-plane orbits (it did not occur for
quadrupole shapes).

\section{Conclusions}

Classical periodic orbits responsible for the emergence of
superdeformed shell structure for single-particle motions in
spheroidal cavities are identified and their relative contributions to
the shell structures are evaluated.  Both prolate and oblate
superdeformations as well as prolate hyperdeformations are
investigated.  Fourier transforms of quantum spectra clearly indicate
that 3D periodic orbits born out of bifurcations of planar orbits in
the equatorial plane become predominant at large prolate deformations,
while butterfly-shaped planar orbits bifurcated from linear orbits
along the minor axis are important at large oblate deformations.

We have also investigated shell structures for reflection-asymmetric
cavities.  It is found that remarkable shell structures emerge for
certain combinations of quadrupole and octupole deformations.  Fourier
transforms of quantum spectra clearly indicate that bifurcations of
triangular and square orbits in the `equatorial' plane play crucial
roles in the formation of these new shell structures.

\end{document}